\documentclass[12pt]{iopart} 
\pdfoutput=1
\usepackage{graphicx} 
\usepackage[caption=false]{subfig}
\usepackage{color}
\usepackage{float}
\usepackage{amsfonts}
\usepackage{booktabs}

\usepackage{lineno}
\usepackage{setspace}

\begin{document} 

\title[]{Growth-rate distributions at stationarity %for stationary systems %processes
}

\author{ Edgardo Brigatti$^{1}$} 
\address{$^{1}$ Instituto de F\'{\i}sica, Universidade Federal do Rio de Janeiro, 
Av. Athos da Silveira Ramos, 149,
Cidade Universit\'aria, 21941-972, Rio de Janeiro, RJ, Brazil}

\ead{edgardo@if.ufrj.br}

%\linenumbers
%\doublespacing

\maketitle

\begin{abstract} 

We propose new analytical tools for describing growth-rate distributions 
generated by stationary time-series.
Our analysis shows how deviations from normality are not 
pathological behaviour, as suggested by some traditional views, but instead
can be accounted for by clean 
and general statistical considerations.
%are accounted for by a specific statistical distribution which can present positive excess kurtosis.
In contrast, strict normality is the effect of specific modelling choices. 

%as suggested in traditional approaches, but the new normal.
%Positive excess kurtosis are not exception but widespread and natural.
%Strictly Normality is an exception of a specific model.

Systems characterized by stationary Gamma or heavy-tailed abundance distributions 
produce log-growth-rate distributions well described by a generalized logistic distribution,
which can describe %generates %shapes who can be close to normality or 
tent-shaped or nearly normal datasets and serves as a useful null model for these observables.
%Pure tent-like shape is connected to error-prone datasets.
These results prove that, for large enough time lags, in practice, growth-rate distributions cease to be time-dependent and exhibit finite variance.
Based on this analysis, we identify some key stylized macroecological patterns and specific stochastic differential equations capable of reproducing them. A pragmatic workflow for heuristic selection between these models is then introduced.
%select a SDE able to reasonably approximate the considered macroecological patterns. 
This approach is particularly useful for systems with limited data-tracking quality, where  
applying sophisticated inference methods is challenging. % cannot be applied.

\end{abstract} 

%{\bf Keywords}:% population dynamics, growth-rate distribution, neutral models, microbiota

%\pacs{89.75.Kd,  05.65.+b, 64.60.-i}

\section{Introduction}

Growth processes are a very widespread phenomenon
which has a prominent relevance in the characterization 
of biological, social and economic systems.
%Traditionally, their description is based on the verification of the Gibrat hypothesis %Gibrat, 1931) of random-walk growth and to consider statistical regularities which are %not described by this approach as a possible violations of a standard considered as a %sort of null hypothesis.
Traditionally, their description is focused on the Gibrat's hypothesis \cite{Gibrat} of 
random-walk growth, which has been considered as a universal starting point 
for the modeling of a variety of systems. % \cite{Holmes}. 
The use of this approach has implied important consequences. 
Gibrat's model generates non-stationary growth processes, 
with abundance time series $x(t)$ characterized by 
%a positive drift in the dynamics of the considered system. 
distributions $P(x)$ approximated by a Lognormal one.
Another important observable used in the description of these dynamics 
is the logarithmic growth rate, defined as: $g(t,\tau) = \ln x(t + \tau) - \ln x(t)$ \cite{Royama,Berryman,Holmes,Brigatti}.
The analysis of its distributions $P(g,\tau)$ 
has generated a great amount of work.
In the case of the classical Gibrat's model $P(g,\tau)$ is normal and, as the 
process is not stationary, its variance grows linearly with time.
%Gherardi_PNAS
This approach is so widespread that it has become a kind of null model for the $P(g,\tau)$ distribution in generic systems, where normality is traditionally assumed based on generic appeals to the central limit theorem.
% where  it %, traditionally, has been considered to be normal, based on generic references to the central limit theorem. 
%Despite this, the fact that many systems have displayed  $P(g,\tau)$ 
%with non-normal, leptokurtic behaviors aroused a lot of interest \cite{}.
However, in the following we will show that for stationary systems, 
it is possible to compute more meaningful distributions to serve as an effective null hypothesis.
These distributions are particularly relevant  as many systems have displayed  $P(g,\tau)$ 
with non-normal, leptokurtic behaviors \cite{Keitt98,Keitt02,Sun15,Ji20,Ashish}.
We will also 
%show how some simple considerations %arising from these observations  
%allow to
 identify key stylized macroecological patterns related to these features and four specific SDEs capable of reproducing them. Based on these results, a pragmatic workflow for heuristic selection between these SDE models will be presented.

In this work, we focus on the case of population dynamics in ecological systems, but our
analyses can be transferred %transliterated 
to other systems and find very broad applications.\\

\section{A general basic setting}
%background} %framework} %Some simple ideas} setting 

Ecological systems are frequently characterized by regulation, which generally 
implies stationarity in abundance time series.
By stationarity we mean that the statistical properties of a time-series do not 
change with a shift in time. 
%are independent of the point in time at which they are observed. 
%In fact, in these series, the joint probability distribution 
%of any subsequence of data points does not change with a shift in time. 
%, eventually affected by seasonality/periodicity. 
For this reason, we move away from the Gibrat's model and we focus on 
models that produce stationary dynamics. 
We consider models based on Markovian stochastic differential equations (SDE), as
they allow for analytical predictions and, in particular, the calculation in closed form 
of the abundance distribution $P(x)$ and, eventually, of the $P(g,\tau)$ \cite{Azaele,Brigatti25}. 

At stationarity, the log-growth rate distribution (LGD) produced by these processes %at this conditions
can be characterized by general properties.
As the generated time series are stationary, the LGD must have all odd-order moments equal to zero.
%This implies that, in order for a parametric distribution  to reproduce these properties, it is enough to have two parameters: a scale (variance) parameter and a shape parameter describing the decay of the tails (kurtosis).
This implies that a distribution  which varies %with two parameters, which control 
in scale (variance) %parameter 
and the shape (kurtosis, which %parameter 
describes the decay of the tails), is,
in principle, enough for successfully approximating its behavior.
Moreover, we can obtain important information %about its functional form 
by calculating the LGD produced by a time series generated by independent random variables drawn from the abundance distribution (AD) generated by a given SDE.
The LGD obtained from this approach, which we denote as $P_\infty(g)$, is interesting for two reasons.

First, it coincides with the $P(g,\tau)$ of the considered SDE model for sufficient large $\tau$ values, when temporal correlations produced by the model cease to be relevant.
%First, it coincides with the limiting distribution $P_\infty(g)$=P(g,\tau>>1)$ of the considered SDE model. 
%This distribution is reached at sufficient large $\tau$ values, when temporal correlations produced by the model cease to be relevant.
In fact, as we are considering Markovian models, correlations  decay exponentially and 
%it follows that they cease to be relevant 
after a finite, characteristic time $t_c$, %after which 
the LGD produced by the model,  i.e. $P(g,\tau>t_c)$, is indistinguishable from %the one calculated with this approach 
$P_\infty(g)$.
%Note that this observation 
This fact implies a non-obvious result which has been previously suggested in \cite{Kalyuzhny}, but not demonstrated: for sufficiently large $\tau$, stationary processes produce an LGD that practically ceases to depend on $\tau$ and displays a finite variance.
 %, and in practice it stops depending on $\tau$.  
%stationary implies finite variance at long enough tau %, as it is well aproximated by $P_\infty(g)$.
Moreover, %as for markovian processes correlations  decay exponentially, 
being Markovian, we expect that  $P_\infty(g)$ is reached in exponential times. In contrast to the case of the Gibrat's model, its variance grows until reaching a saturation value corresponding to the variance of $P_\infty(g)$, following an exponential law:
$Var[P(g,\tau)] \approx Var[P_\infty(g)]-a_1e^{-\frac{\tau}{b_1}}$,
%(1-ae^{-\frac{\tau}{b}})Var[P_\infty(g)]$,
and similarly, its kurtosis follows:
$Kur[P(g,\tau)] \approx  Kur[P_\infty(g)]+a_2e^{-\frac{\tau}{b_2}}$.%ae^{-\frac{\tau}{b}}+kur[P_\infty(g)]$.

Second, $P_\infty(g)$ can be used as a reasonable null hypothesis when rigorously testing the relevance of a specific SDE in describing $P(g,\tau)$. This null distribution corresponds to a general process presenting the expected AD but  without the peculiar temporal correlations produced by a specific SDE. In fact, this distribution is generated by the basic statistics of the AD and it is not determined by the specific form of the chosen SDE.\\

\section{Implementation for modeling and forecasting}

We can implement these ideas starting from some typical empirical patterns. 
%For example, a first situation corresponds to 
We can consider empirical ADs well approximated by a Gamma distribution: $\Gamma(\alpha,\beta)$. 
This is  a common situation in ecology \cite{Brigatti25,Grilli,Engen96G}, as these distributions are very flexible %statistical models for 
and can describe very different non-negative datasets. 
%On varying its parameters Erlang, chi-squared and exponential
Numerous distributions can be seen as special cases of the Gamma . %are members the gamma family, 
Moreover, for a vanishing shape parameter it approximates the Fisher distribution, for large ones
converges to a normal one, and it can describe datasets close to log-normality.
For a Gamma distribution,  $P_\infty(g)$ can be calculated from the ratio of two independent Gamma random variables,
once the logarithmic change of variable is realized, %x=log(y), 
obtaining:
\begin{equation}
P^{\Gamma}_\infty(g,\alpha)=\frac{\Gamma(2\alpha)}{\Gamma(\alpha)^2} \frac{e^{\alpha g}}{(e^{g}+1)^{2\alpha}}, 
\label{eq1}
\end{equation}
which is a generalized  logistic distribution (for details, see \cite{SuppMat}).
%that for $\alpha = 1$, it is a logistic distribution, a well known distribution with a shape quite similar to the normal one, but with heavier tails.
By varying its parameter, different values of kurtosis and variance can be produced. For small $\alpha$ it presents large kurtosis (leptokurtic), while for $\alpha>3$ it is easy to confuse it with a normal distribution ($N(0,\sigma^2)$)\cite{SuppMat}.
%becomes practically indistinguishable from a normal distribution ($N(0,\sigma)$)\cite{SuppMat}. 
Although the values of the two moments are dependent on each other,  this distribution can reasonably fit very different datasets. 
Note that its tails are very close to those of a Laplace distribution \cite{SuppMat}. 

A Gamma AD can be produced 
%not only by a logistic model with environmental noise, as already reported in \cite{Ashish}, but also 
by any SDE belonging to the %generalized 
family of  equations given by:
%$dx_t= (b-ax_t)dt+\sqrt{2D x_t}dW_t$.
%$\frac{dx}{dt}= x^{k-1}(a-bx)+\sqrt{2c} x^{k/2} \eta(t)$, 
$dx_t= [x_t^{k-1}(a-bx_t)]dt+\sqrt{2c} x_t^{k/2} dW_t$, where $W_t$ is a standard Wiener process,
$k \in \mathbb{R}$, %\setminus \{0\}$.  
with positive parameters and $a/c>k-1$ (for details, see \cite{SuppMat}). 
Among these models, the cases with $k=1,2$ have 
%a compelling qualitative rationale.
been frequently used to model ecological systems:
$k=1$ corresponds to a biogeographic model which includes migration and is drift-dominated \cite{Azaele,Brigatti25}, 
and $k=2$ is the standard stochastic logistic model \cite{eLife,Grilli}, which neglects migration and presents explicit environmental noise.
% $k=0$??  .
For $k=1$  it is possible to obtain the LGR in a closed form, which we denote with $P^*(g,\tau)$ \cite{Azaele,SuppMat}, and for $k=2$ in an approximate form \cite{Brigatti25}. These analytical results
are consistent with the general predictions on the LGR behaviors made in section 2 (details in \cite{SuppMat}). 
 
The distribution of eq. \ref{eq1} can also be produced considering an AD 
well approximated by an Inverse-Gamma distribution.
These distributions are of particular interest as they 
present  tails with a power-law and an exponential cutoff and
can provide a good approximation for population characterized by heavy tails. 
%These ADs can be produced
They can be generated by any SDE belonging to the %generalized 
family of  equations given by:
%$dx_t= (b-ax_t)dt+\sqrt{2D x_t}dW_t$.
%$\frac{dx}{dt}= x^{2-k}(a-bx)+\sqrt{2c} x^{2-k/2} \eta(t)$, 
$dx_t= [x_t^{2-k}(a-bx_t)]dt+\sqrt{2c} dx_t^{2-k/2} dW_t$, with
$k \in \mathbb{R}$, %\setminus \{0\}$.  
positive parameters and $b/c>k-3$ \cite{SuppMat}. 
The case with $k=2$ corresponds to a linear drift term with an environmental noise,  and presents a compelling qualitative rationale.

Another interesting situation is an AD presenting a Lognormal distribution \cite{Engen96,Engen02}. In this case, the $P_\infty(g)$ obtained by our approach is normal. 
This distribution can be generated by a SDE
with an environmental noise $\sqrt{2c} x_t dW_t$ and %a cumbersome 
a drift %term
%$ax-bx \ln(x) +\frac{c}{x}$ \cite{SuppMat} il problema e' il 1/x, il resto e'  gompertz...  
$(a-c)x_t-bx_t \ln(x_t)$ \cite{SuppMat}, which presents a Gompertz-type growth %density dependent
term.
These results are summarized in Table \ref{Tab}.

Finally, for uniform AD, following the same procedure we obtain a $P_\infty(g)$ with a Laplace distribution  with zero location and unitary scale parameter (see \cite{SuppMat}).
This AD corresponds to an abundance distribution strongly affected by sampling bias, which is entirely determined by the detection probability.
This result, together with the fact that the asymptotic behavior of the tails of $P^{\Gamma}_\infty(g,\alpha)$ decay exponentially, can explain why the Laplace distribution can reasonably fit some dataset and why it has been heuristically used in numerous previous works \cite{Ji20,Sun15,Stanley,Plerou99}.

% for $g \to \pm \infty$ and fixed $\tau$, follow $e^{-\alpha g}$. also the other scale as exp!!!!}

\begin{table}[h]
\centering
\small
\begin{tabular}{lccccc}
\toprule
SDE & SDE & Abundance  & $P_\infty(g)$ & $P(g,\tau)$ \\
Type &  &  Distribution &  & \\
\midrule
I & $dx_t= (a-bx_t)dt+\sqrt{2c x_t}dW_t$ & $\Gamma(\alpha,\beta)$ & $P^\Gamma_\infty(g,\tau)$ & $P^*(g,\tau)$   \\
II & $dx_t= x_t(a-bx_t)dt+\sqrt{2c} x_tdW_t$ & $\Gamma(\alpha,\beta)$ & $P^\Gamma_\infty(g,\tau)$ & $\sim N(0,\sigma^2(\tau))$   \\
III & $dx_t= (a-bx_t)dt+\sqrt{2c} x_t dW_t$  & Inv-$\Gamma(\alpha,\beta)$ & $P^\Gamma_\infty(g,\tau)$ & ---   \\
%IV & $dx_t= (ax_t-bx_tlog(x_t)+c/x_t)dt+\sqrt{2c} x_t dW_t$ & Lognormal & $N(0,\sigma^2)$ & ---  \\
IV & $dx_t= [(a-c)x_t-bx_t \ln(x_t)]dt+\sqrt{2c} x_t dW_t$ & Lognormal & $N(0,\sigma^2)$ & ---  \\
\bottomrule
\end{tabular}
\caption{The four SDEs considered, along with the macroecological patterns each produces.}
\label{Tab}
\end{table}

\section{Results and Discussion}

%From this simple reasoning it is possible to draw some important conclusions.
The description of the $P(g,\tau)$ of stationary processes with the normal distribution is not
the norm, %, as it has been suggested by the literature, 
but it is produced by specific SDE models.
In particular, we know that for the logistic model with environmental noise 
$P(g,\tau)$ is well approximated by a normal distribution for every $\tau$, 
but must converge in the limit  $\tau\to\infty$ to $P^\Gamma(g,\alpha)$ \cite{SuppMat}.
$P_\infty(g)$ is strictly normal only for the SDE which produces Lognormal AD.
 
Non-normal, leptokurtic behaviors in the LGD are not an exception, and are commonly found in empirical data. 
They can be easily justified by our statistical reasoning, from which the correct null-hypotheses, for different ADs, are obtained. First, the generalized logistic distribution of  eq.\ref{eq1} for Gamma and Inverse-Gamma AD. 
This null distribution corresponds to a process presenting the expected AD but without the peculiar temporal correlations produced by a specific SDE.
Second, the Laplace distribution for AD corresponding to dataset overwhelmed by noise.\\
%In the case the best fitted LGD is a Lapalacian distribution we can assert that the dataset is overwhelmed by noise.

We can define a workflow for realizing a practical heuristic SDE model selection, based on the following stylized
macroecological patterns defined by our approach: time evolution of $Var[P(g,\tau)]$, AD, $P(g,\tau=1)$ and $P(g,\tau>>1)$ distributions.
Saturating behaviors in $Var[P(g,\tau)]$ suggest that the process is Markovian.  For an AD with a Lognormal distribution and  $P(g,\tau>>1)\sim N(0,\sigma^2)$ the process can be approximated by the Type IV SDE,  for an AD with an Inverse-Gamma  distribution and 
$P(g,\tau>>1)\sim P^{\Gamma}_\infty(g,\alpha)$  by the Type III SDE, for an AD Gamma distributed and 
$P(g,\tau>>1)\sim P^{\Gamma}_\infty(g,\alpha)$, if $P(g,\tau=1)\sim N(0,\sigma^2)$ the SDE can be the Type II,
if $P(g,\tau=1)$ is approximated by $P^*(g,\tau)$, the Type I model. 
This approach is rudimentary and generally relies on sufficient conditions, but can select a SDE able to reasonably approximate the considered macroecological patterns.  This classification framework is summarized in Fig. 1.
It is particularly useful for systems, like the ecological ones, with limited quality %availability 
in tracing data (scarce, low frequencies, noisy data), where standard, sophisticated inference methods \cite{Iacus} can not be applied.

\begin{figure}[h]
\begin{center}
\includegraphics[angle=0,width=1.\textwidth]{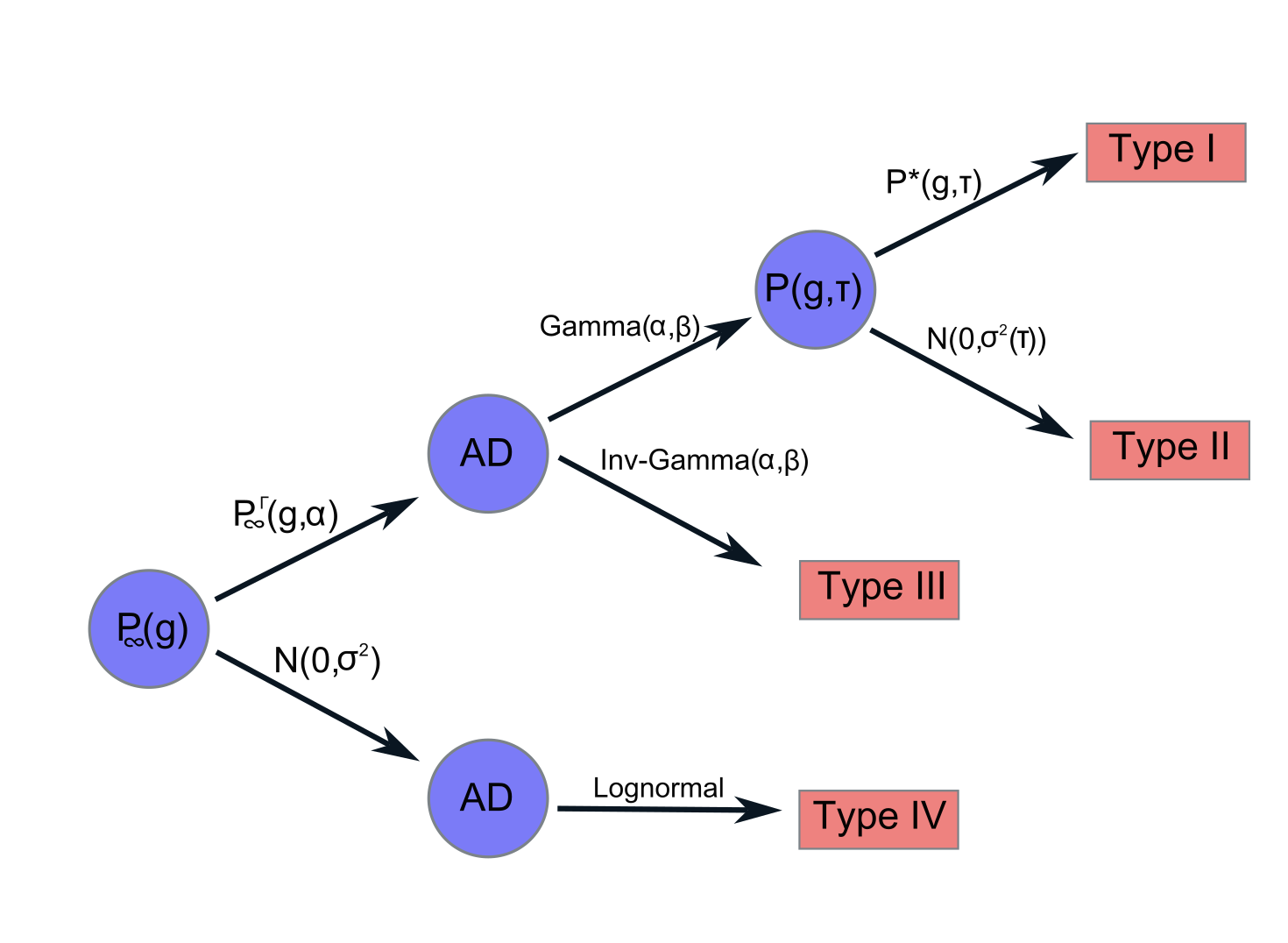}
\end{center}
\caption{\small A decision tree illustrating our classification framework.
}
\label{Fig_tree}
\end{figure}

%With the limited availability of ecological tracing data, it is challenging to apply standard inference methods
For more rigorous approaches, we must warn that, when available, it is crucial to analyze properties that explicitly depend on more specific temporal correlations generated by the SDE, such as the temporal evolution of the LGD in dependence of $\tau$ \cite{Brigatti25}. 
By including these temporal patterns is
%by including these %substantial statistical analyses it is 
possible to obtain more robust results, given that in the AD the temporal structure is absent and in the $P(g,\tau)$ it can be difficult to detect. 
%It follows that given an abundance time-series described by these distributions, departure from normality in the LGD are not an hint in favor of specific SDE. 
Moreover, general departures from normality in the LGD cannot be used as a sufficient condition in favor of a specific SDE. 
In particular, in the case of Type I SDE, $P^{\Gamma}_\infty(g,\alpha)$, and not a generic normal distribution, must be used as the correct
%For this reason, these 
null hypothesis when fitting $P(g,\tau)$ with $P^*(g,\tau)$. % for rigorously selecting the correct SDE.

\section{An application to empirical data}
 
%We show the relevance of the considered macroecological patterns and 
We apply our framework to selected abundance time series from the Global Population Dynamics Database \cite{GPDD}. 
The dataset was restricted to series presenting a sufficient length, 
modest seasonality effects and stationarity (see \cite{SuppMat} for details).
%It is interesting to note that less than 30\% of time-series satisfies these constraints (60/203), showing that effectively stationary and clearly not periodic time series are a common outcome of population dynamics.

We clearly found the expected behaviour of $Var[P(g,\tau)]$ in 78\% of the considered series;
43\% of the data present the exponential growth towards a finite value of the variance and
35\% show a  constant value, suggesting that the sampling time is larger than
the correlation time of the process.
The remaining series present a linear growth of the variance. Assuming that the stationarity tests are correct, we can imagine that the system is endowed with very strong correlations produced by memory effects (the Markovian approximation is not valid) or the variance is stuck in the growing phase without reaching the saturation value (very slow dynamics). 

%In the following we report results 
%from the series presenting the expected behaviour for the LGR variance, but similar results can be obtained 
%considering all the analysed series.
%The analysis shows the relevance of the considered macroecological patterns.
The ADs are best described by a Gamma  (47\% of series) or a Lognormal distribution (47\%) %of the series 
and the remaining series are characterized by an Inverse-Gamma.
$P(g,\tau=1)$ is best fitted by $P^*(g,\tau)$ in 73\% of the cases, and the remaining 27\% by $N(0,\sigma^2)$.
%For the remaining case it is not possible to distinguish between the two distribution.
The $P(g,\tau>>1)$ generally presents a small excess kurtosis and 30\% of the case
are best fitted by $P^{\Gamma}_\infty(g,\alpha)$ and the remaining by $N(0,\sigma^2)$.
Perhaps there is an overestimation of normal distributions, as for large $\alpha$ it can be difficult to distinguish between them. Examples of this phenomenology can be found in Fig. 2.
 
Using the workflow summarized in Fig. 1, we can coherently classify 73\% of the time series:
%We obtain that  
25\% of the time series are approximated by Type I SDE, 30\% by Type II, 7\% by Type III and 38\% by Type IV (see \cite{SuppMat} for more details).

Finally, for the Type I SDE, we check the consistency of the different methods used to infer the parameter 
values. 
We compare the three independently estimated values of $\alpha$, which we indicate as 
$\hat{\alpha}_{\tau=1}$ (obtained from $P(g,\tau=1)$), $\hat{\alpha}_{\tau>>1}$ (obtained from$P(g,\tau>>1)$) and $\hat{\alpha}_{Ab}$ (obtained from the AD). Results are shown in Fig. 3.
The correlation between $\hat{\alpha}_{\tau>>1}$ and $\hat{\alpha}_{Ab}$ is impressive.
The correlations of  $\hat{\alpha}_{\tau=1}$ and $\hat{\alpha}_{\tau>>1}$ with $\hat{\alpha}_{Ab}$ present a higher dispersion of the points. 
Taking into account the size and noise of the time-series, the result suggests that the different approaches
used to estimate the parameters are consistent. 

In conclusion, empirical data show the relevance of the considered macroecological patterns: Gamma and Lognormal distributions can well approximate the AD distribution and, more rarely, also the Inverse-Gamma. The LGR are well described by the considered analytical distributions, for both short and long $\tau$ lags, and the expected behavior of their variance is confirmed by empirical data. The workflow can successfully select the SDE that reasonably reproduces the patterns considered. 
These SDEs can be associated with specific underlying mechanisms responsible for the observed phenomenology of these population dynamics.

\begin{figure}[h]
\begin{center}
\includegraphics[angle=0,width=1.\textwidth]{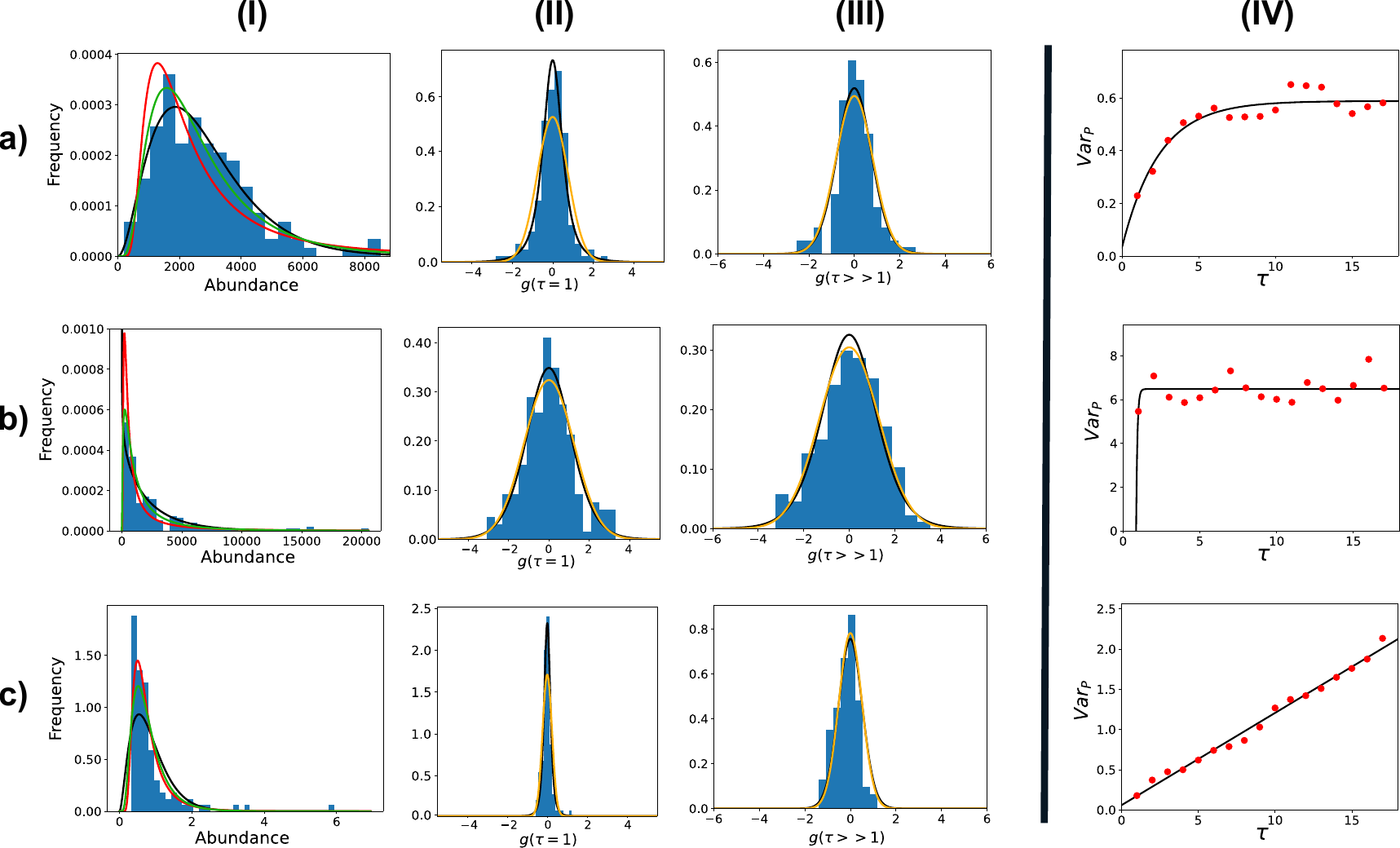}
\end{center}
\caption{\small On the left: Columns correspond to the abundance distribution (I), the LGR distribution at $\tau=1$ (II), and the LGR distribution at $\tau>>1$ (III) for typical time-series with abundance distribution best fitted by a Gamma (a), a Lognormal (b) and an Inverse (c) distribution.
The solid lines represent the fits: Gamma (black), Lognormal (green), and Inverse (red).
In column II the solid black lines represent the fits with $P^*(g,\tau)$, in column III 
the fits with $P_\infty(g)$. Yellow lines are adjusted using a Normal distribution.
On the right: column IV shows typical examples of the different behaviours of the evolution of $Var_P=Var[P(g,\tau)]$ as a function of $\tau$.
For more details, see \cite{SuppMat}.
}
\label{Fig_dynamics}
\end{figure}

\begin{figure}[h]
\begin{center}
\includegraphics[angle=0,width=0.49\textwidth]{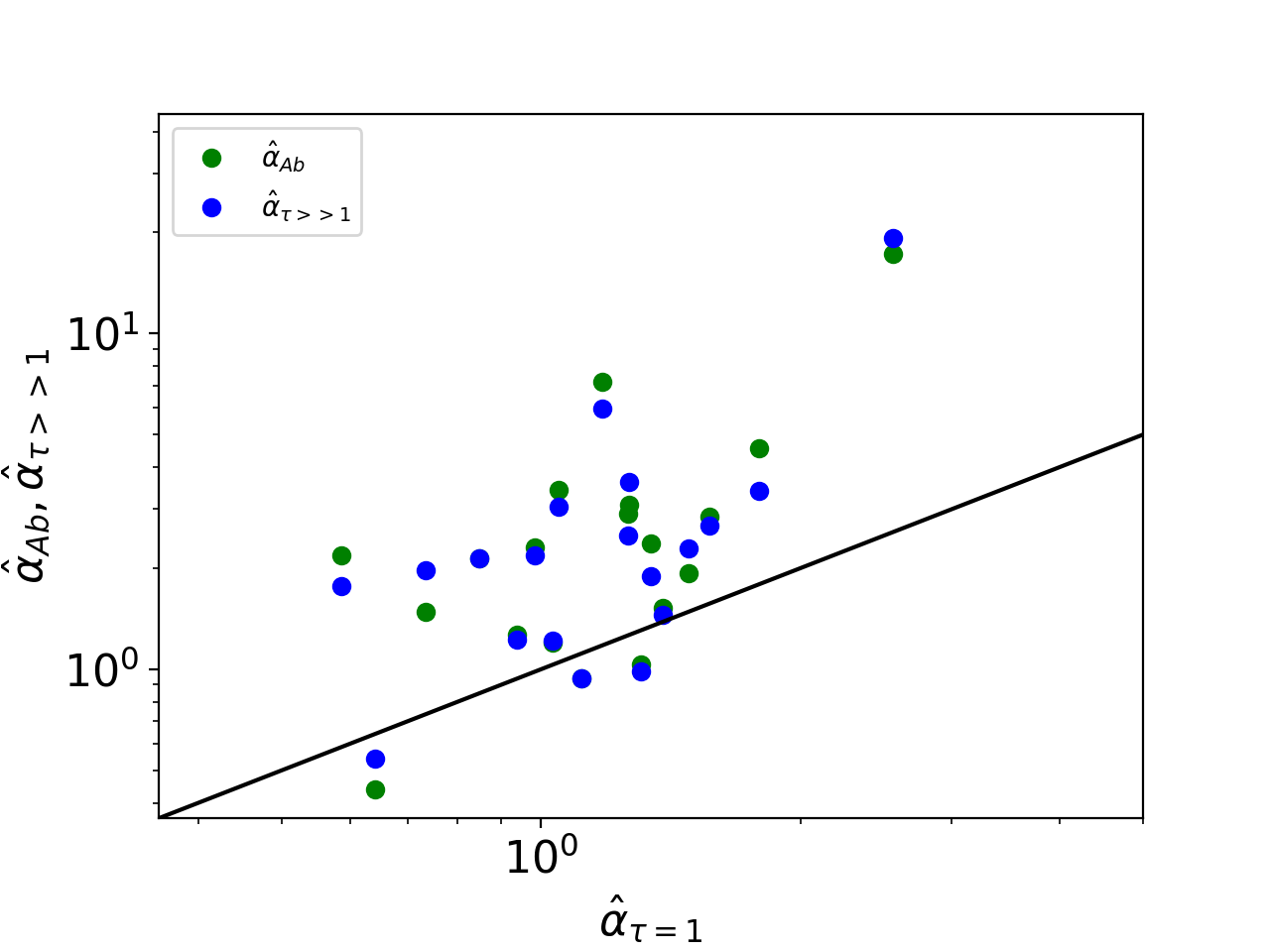}
\includegraphics[angle=0,width=0.49\textwidth]{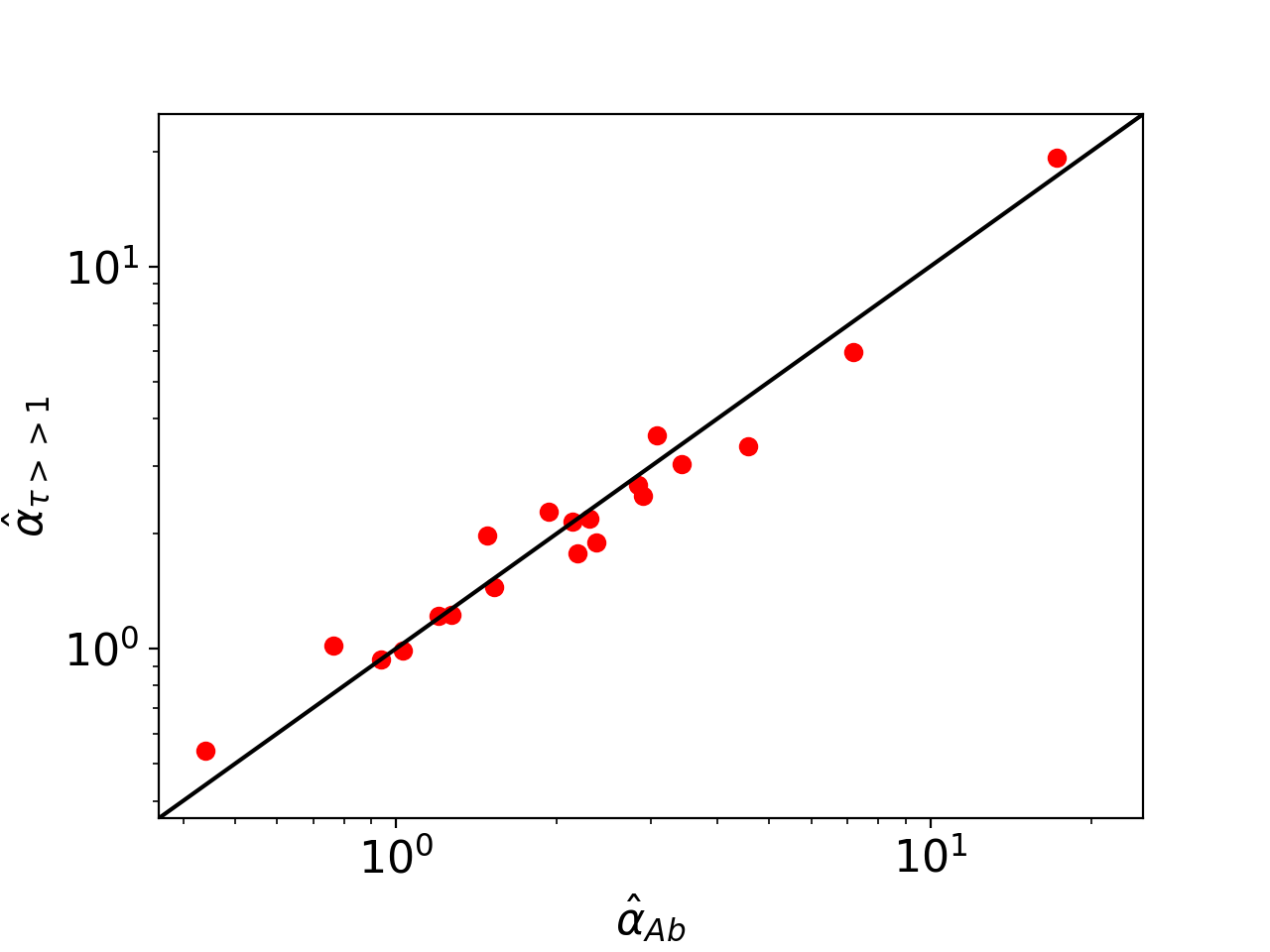}
\end{center}
\caption{\small On the left: In this scatter plot blue points represent
the $\hat{\alpha}_{\tau=1}$ %($\alpha$ estimated from $P(g,\tau=1)$),
versus $\hat{\alpha}_{\tau>>1}$. 
%($\alpha$ inferred from $P(g,\tau>>1)$)
Green points stand for $\hat{\alpha}_{\tau=1}$ versus $\hat{\alpha}_{Ab}$.
%($\alpha$ estimated from the stationary abundance distribution).
The Pearson correlation is equal to 0.73 and 0.76, respectively. %to 0.77 and 0.81 witthout growing Var_p
On the right: Red points stand for $\hat{\alpha}_{\tau>>1}$ versus $\hat{\alpha}_{Ab}$.
The Pearson correlation is equal to 0.98.
The solid lines are $x=y$.
%The mean value of $\hat{\alpha}_{\tau=1}$ is equal to $1.45\pm0.41$, which is consistent with the value found analyzing the ensemble of all time-series together.
%$b/D$ values estimated from the log-growth distribution for  $\tau=1$ versus the estimations for $\tau>>1$ (blue points).Red points represent $b/D$ values estimated from the log-growth distribution for  $\tau=1$ versus the ones estimated from the abundance stationary distribution.
%X: $b/D$ estimated from P(g,\tau=1), Y: b/D measured from long
%X: b/D measured from P(g,\tau=1), Y: b/D measured from stationary (red points).
}
\label{Fig_scatter}
\end{figure}

\section*{Acknowledgments}

The author received partial financial support from the National Council for Scientific and Technological Development - CNPq, Grant No. 305984/2024-1.

%Data availability statement: This paper does not use original data. All datasets used in the paper are available from the original references. The code used in the analysis of these datasets is available at https://zenodo.org/ badge/latestdoi/400120651.

\end{document}